# Bound excitons and bandgap engineering in violet phosphorus

Zhenyu Sun,[1,2] Zhihao Cai,[1,2] Peng Cheng,[1,2] Lan Chen,[1,2,3] Xuewen Zhao,[4] Jinying Zhang,[4] Kehui Wu,[1,2,3,5*] Baojie Feng,[1,2,5*]

[1]*Institute of Physics, Chinese Academy of Sciences, Beijing, 100190, China*
[2]*School of Physical Sciences, University of Chinese Academy of Sciences, Beijing, 100049, China*
[3]*Songshan Lake Materials Laboratory, Dongguan, Guangdong, 523808, China*
[4]*State Key Laboratory of Electrical Insulation and Power Equipment, Center of Nanomaterials for Renewable Energy, School of Electrical Engineering, Xi'an Jiaotong University, Xi'an, 710049, China*
[5]*Interdisciplinary Institute of Light-Element Quantum Materials and Research Center for Light-Element Advanced Materials, Peking University, Beijing, 100871, China*

*Corresponding author. E-mail: khwu@iphy.ac.cn; bjfeng@iphy.ac.cn.

**Keywords:** violet phosphorus, bound exciton, bandgap engineering, strain, 2D semiconductor

## Abstract

Violet phosphorus (VP), the most stable phosphorus allotrope, is a van der Waals semiconductor that can be used to construct *p*-type nanodevices. Recently, high-quality VP crystals have been synthesized while a deep insight into their excitonic properties and bandgap tailoring approaches, which are crucial for their optoelectronic device applications, is still lacking. Here, we study the optical properties of ultrathin VP by second harmonic generation, photoluminescence, and optical absorption spectroscopy. We observed strong bound exciton emission that is 0.48 eV away from the free exciton emission, which is among the largest in 2D materials. In addition, the bandgaps of VP are highly sensitive to the number of layers and external strain, which provides convenient approaches for bandgap engineering. The strong bound exciton emission and tunable bandgaps make VP a promising material in optoelectronic devices.

**Introduction**

Two-dimensional (2D) semiconductors, exemplified by transition metal dichalcogenides and black phosphorus, have attracted great interest because of their potential applications in next-generation quantum devices [1-5]. Their optical responses vary from mid-infrared [4] to deep ultraviolet [5], which enables optoelectronic device applications over a wide range of wavelengths. As the most stable phosphorus allotrope, violet phosphorus (VP) is a van der Waals material and is expected to host hole mobility of up to 7000 $cm^2V^{-1}s^{-1}$ and a direct bandgap of 2.5 eV in the monolayer limit [6]. In 2020, high-quality VP single crystals were synthesized [7-9], which has stimulated intensive research interest in this emerging material. Later, a series of novel properties were discovered in VP, including impurity-induced trionic effects [10], in-plane anisotropic phonons [11], VP-based photodetection [10], and VP-embedded complementary metal oxide semiconductor (CMOS) inverter arrays [12].

In some 2D semiconductors, bound exciton emission induced by in-gap states, together with free exciton emission originating from the optical bandgap transitions, have dominant influences on their optical responses [13,14]. In addition, bound excitonic emissions can behave as single quantum emitters, which are essential for applications in quantum information processing [15]. However, bound excitons are usually very close to free excitons (<300 meV) due to the shallow energy levels of localized in-gap states from the bottom of the conduction band [16,17]. From this viewpoint, searching bound excitons well separated from the free excitons is highly desirable to avoid interplay among different emission channels. In VP, previous photoluminescence (PL) measurements revealed the existence of both neutral and charged excitons [10], while studies on the possible bound excitons are still lacking.

Besides excitonic properties, bandgap engineering is also of great significance for 2D semiconductors because of its potential use in manipulating optical and electronic properties [18]. To date, several approaches have been proposed to modulate the bandgaps of 2D semiconductors. For example, the bandgap of many materials depends on their thicknesses because of the quantum confinement effects along the stacking direction [19-21]. In addition, applying external strain is another effective method to manipulate the bandgaps of 2D semiconductors [22,23]. Due to the layered structure and relative high elasticity/Young's modulus, their electronic structures are prone to be influenced by external strain. For monolayer $MoS_2$, a prototypical 2D semiconductor, a bandgap modulation up to ~300 meV and a modulation rate of ~136 meV/% were achieved by applying uniaxial strain [23].

In this work, we study the optical properties VP by second harmonic generation (SHG), PL, and optical absorption spectroscopy. We observed an unexpected SHG signal in VP, which is quite rare in 2D centrosymmetric semiconductors. Our temperature-dependent PL measurements revealed strong bound exciton emission at 1.43 eV, besides the free exciton emission at 1.91 eV. In addition, the bandgap of VP is tunable by controlling the number of layers and applying tensile strain. When the VP flakes are thinned down from the bulk to the bilayer, their bandgaps increase from 1.78 to 2.23 eV. Finally, we found that applying tensile strain can change the bandgap by 100 meV.

**Results**

**Bound excitons in violet phosphorus**

Single-crystal VP is a van der Waals material with a monoclinic structure (space group: P2/n). It is stacked along the *c*-axis with an interlayer spacing of ~1.1 nm [8], as shown in Fig. 1a. Each unit cell of VP has 84 atoms with various bonds, giving rise to multiple vibrational modes, as evidenced by our Raman spectroscopy measurements (Fig. 1b). Intense Raman peaks that derive from the breathing and stretching modes of the phosphorus cages are observed at 353, 358, 373, and 471 cm$^{-1}$, in agreement with previous reports [24]. Due to the weak van der Waals interactions between adjacent layers, VP can be easily thinned down to a few and even monolayers by mechanical exfoliation, as shown in Fig. 1c and 1d.

Figure 1e shows the SHG measurement results on a VP flake (see Supplementary Fig. 1) using 1064-nm excitation, and surprisingly, a pronounced SHG signal was observed. Typically, the SHG signal indicates the broken inversion symmetry. Since bulk VP has a centrosymmetric crystal structure, the SHG signal might originate from the topmost odd layers that lacks inversion symmetry. Similar phenomena have been reported in other monoclinic crystals [25]. Moreover, our polarization-resolved measurements show that the SHG signal is anisotropic with the polarization extinction ratio (PER) in the parallel and perpendicular configurations being 3.4 and 2.5, respectively, as shown in the inset of Fig. 1e and Supplementary Fig. 2. By varying the excitation laser power, we observed a linear relationship between the SHG intensity and the excitation power in the log-log plot (see Supplementary Fig. 3). The fitted slope 1.8 is close to the theoretical value 2, further confirming the second-order nonlinear process nature [26].

We then study the excitonic properties of VP by temperature-dependent PL measurements. The PL signal corresponds to the allowed optical transitions between occupied and unoccupied states under an external photon stimulus. At low temperatures, in-gap states, which might originate from defects or dislocations, will have a higher contribution to the PL signal [27]; this is because the thermal activation energy at low temperatures is relatively small and inadequate to activate the free excitons trapped by localized sites [28].

A typical PL spectrum taken at 5 K with 532-nm excitation is displayed in Fig. 2a. The peak at ~1.91 eV (denoted as $X_G$) originates from free excitons, which corresponds to the optical bandgap of VP. Interestingly, we observed an additional peak at ~1.43 eV (denoted as $X_B$), which has not been reported before. The intensity of $X_B$ reduces as the VP flake gets thinner, but the line shape and peak position almost keep invariant (see Supplementary Fig. 4). However, when the wavelength of the incident laser is switched to 633 nm, only $X_B$ can be observed with a negligible peak position shift, which suggests an excitonic emission related to the in-gap states instead of an electronic Raman scattering process. The suppression of $X_G$ with 633-nm photons is reasonable because the excitation photon energy (1.96 eV for 633 nm) is very close to the optical bandgap of VP and is difficult to drive sufficient electrons to free exciton states [8]. On the other hand, 1.96-eV photons are high enough to populate electrons of in-gap bound states, and electron-hole recombination from these states to the valence bands leads to a strong PL signal. Notably, the energy separation of $X_B$ and $X_G$ is as large as ~0.48 eV, much larger than that of most 2D semiconductors [29-34]. Such significant energy separation makes it possible to access individual states without interference, which is feasible for practical applications in optoelectronic devices.

To better understand the origins of the two PL peaks, we performed excitation-power-dependent PL measurements at 5 K ($\lambda_{ex}$=532 nm), and the results are shown in Fig. 2b. The maximum PL intensity (*I*) as a function of excitation power density (*P*) is displayed in Fig. 2c. We found that both emission peaks follow the power law relationship $I \propto P^k$. The fitted exponent *k* of $X_G$ is ~0.94, indicating a linear relationship with the excitation power. Such linear dependence is a character of direct recombination of free excitons [35]. In contrast, the fitted exponent $k$ of $X_B$ is ~0.58, close to that of bound excitons in other materials [36,37]. When the irradiation power is high enough, the intensity of $X_B$ will saturate because of the nearly full population of the bound states [38].

Figure 2d and 2e show temperature-dependent PL spectra from 20 K to 380 K. As the temperature increases, the intensities of both $X_B$ and $X_G$ decrease. Notably, $X_B$ almost disappears at temperatures above 180 K with a dramatic thermal broadening (see Supplementary Fig. 6). On the other hand, $X_G$ is still observable up to 380 K while the PL linewidth changes a little (see Supplementary Fig. 6). Such thermal dependence behavior of $X_B$ indicates a localized-state-induced excitonic emission rather than a strain effect because the strain-induced excitonic emission is not sensitive to thermal activation [31-34]. The intensity of $X_B$ as a function of temperature is shown in Fig. 2f. At low temperatures, abundant electrons are trapped at localized in-gap states after optical excitation, forming $X_B$ that radiatively decays to the valence band. With increasing temperature, trapped electrons can be thermally activated into delocalized states and captured by nonradiative decay channels [37], which can explain the reduction of the intensity of $X_B$. A schematical diagram is shown in Supplementary Fig. 7. The temperature-dependent intensity can be described by a thermal dissociation process [39]:

$$I(T) = \frac{I(0)}{1 + \left(\frac{\tau}{\tau_0}\right) e^{-\frac{E_A}{kT}}} \tag{1}$$

where $I(0)$ is the PL intensity at 0 K, $E_A$ the activation energy representing the thermal fluctuation energy to dissociate trapped excitons, $\tau$ the excitonic lifetime, and $\tau_0$ the effective scattering time. By fitting the data in Fig. 2f with Equation 1, the activation energy ($X_B$) is approximately 29.5 meV, which is comparable to other 2D materials such as monolayer $WSe_2$ [38,39].

A common origin of the bound excitons in 2D materials is defects that might trap free excitons at low temperature [27,29]. To clarify of the origin of the bound excitons in VP, we performed a detailed element analysis on freshly exfoliated VP flakes and found that the VP flakes are highly pure, as shown in Supplementary Fig. 8. Therefore, the bound exciton emission in VP is likely to originate from phosphorus-related defects or the special bi-tubular structure of VP.

**Layer-dependent bandgap evolution**

Next, we explore the strategies of bandgap engineering of layered VP. First, we studied the bandgap evolution of VP as a function of the number of layers. Microscopic optical absorption spectroscopy is a powerful tool to study the optical bandgaps of 2D materials, typically performed in the reflectance or transmission mode. Although signatures of thickness-dependent bandgaps have been reported in liquid-phase exfoliated VP samples [9], VP flakes in solvents after centrifugation usually have a wide thickness distribution, and a detailed study on the layer-dependent bandgaps is still lacking.

To this end, we prepared a series of ultrathin VP crystals down to bilayers (see Supplementary Fig. 9) on sapphire substrates by mechanical exfoliation. Detailed optical absorption measurements were carried out on these samples by comparing the reflected light with the excitation light. The absorbance is proportional to $\Delta R/R$, considering the thickness of VP is much smaller than the wavelength of the excitation light [40]. With decreasing number of layers, the absorption intensity gradually decreases, accompanied by an apparent blue shift of the absorption edge, as shown in Fig. 3a. The bandgaps of VP can be extracted from the spectra using the Tauc plots [41], and the results are summarized in Fig. 3b. For an indirect-gap semiconductor, the Tauc formula gives that:

$$(Ah\nu)^{\frac{1}{2}} = B(h\nu - E_g) \qquad (2)$$

where *A* is the measured absorption, *B* a constant, and $E_g$ the optical bandgap. By extrapolating the linear region near the optical absorption edge, we can determine the bandgaps from the intersection points on the horizontal axis. Figure 3c shows the optical bandgaps of VP versus the number of layers. One can see that the gap size increases monotonically from 1.78 to 2.23 eV with decreasing thicknesses. The bandgap increase in thinner samples is caused by the quantum confinement effect, which has also been reported in other 2D semiconductors [20,21].

**Tuning the bandgap by tensile strain**

Besides the number of layers, external strain also strongly influences the bandgap of VP. Recently, various methods have been proposed to apply strain on van der Waals crystals. By optical or scanning tunneling spectroscopy measurements, bandgap tailoring was realized in many 2D materials [42,43]. In these methods, a common approach is to apply uniaxial strain by a two-point bending apparatus, and the flakes are placed on a flexible substrate [44]. However, this method is often ineffective for thicker samples [45] because of strain release in the topmost layers (see Supplementary Fig. 10). To realize effective strain transfer to thick VP flakes, which have a stronger PL response, we use pre-patterned circular arrays and place the exfoliated VP flakes on the as-prepared substrates by all-dry viscoelastic stamping. In this way, the VP flakes will be naturally deformed by the protuberant nanostructure as a whole, regardless of the thickness.

After standard UV lithography and metal lift-off procedures, a large-area Au array was fabricated, as shown in Fig. 4a. When VP nanosheets are transferred onto the substrate, their shapes keep intact in flat areas. However, near the edges of the Au disks, the VP flakes have a strong mechanical strain, which might change the bandgap. The black and red curves in Fig. 4b shows the PL spectra taken on flat and strained areas, respectively. The PL peak of the strained area has an apparent redshift from 1.88 to 1.82 eV compared to that of the unstrained area, which indicates a strain-induced decrease of the bandgap, whereas the Raman peak position shows negligible shift (see Supplementary Fig. 11). The bandgap reduction might originate from the reduced orbital hybridization and bandwidth due to the increased bond lengths [22]. Moreover, we observed a similar redshift for the bound exciton emission, as shown in Supplementary Fig. 12. Notably, both $X_B$ and $X_G$ show no evident polarization dependence (see Supplementary Fig. 5 and Fig. 13).

To further demonstrate the strain-induced effects on the bandgap, we performed PL mapping on a continuous VP flake on top of an Au disk, as shown in Fig. 4c and 4d. At 1.82 eV, we observed a ring-like feature corresponding to the periphery of the Au disk. The enhanced PL intensity at the strained area directly shows the strain-induced bandgap reduction. To better visualize the shift of the PL peak in the strained area, we presented normalized PL spectra across the Au disk, as shown in Fig. 4e. We found that the maximum bandgap shift is ~100 meV accompanied by an apparent PL linewidth broadening in the strained area (see Supplementary Fig. 14). Similar results have been reported in other 2D materials under tensile strain [22,23,44].

**Discussion**

To summarize, we performed a systematic optical study on VP. We observe SHG signals with two-fold symmetry in VP flakes. Our temperature-dependent PL spectroscopy measurements reveal a strong bound exciton emission at 1.43 eV below 180 K, which is well separated from the optical bandgap. In addition, the bandgaps of VP flakes are highly sensitive to the thickness and external strain, which provides a convenient approach for bandgap engineering. The strong bound exciton emission and tunability of bandgaps call for further efforts to fabricate VP-based optoelectronic devices in the future.

**Methods**

**Sample preparation.** High-quality VP crystals were synthesized by a chemical vapor transport (CVT) method. VP nanoflakes were obtained following the standard mechanical exfoliation procedure using blue Nitto tapes. Exfoliated flakes can be transferred onto any substrate such as $SiO_2$/Si, sapphire, and pre-patterned array. To transfer VP flakes onto pre-patterned arrays, we pressed a viscoelastic polydimethylsiloxane (PDMS) stamp covered with thick VP flakes onto the substrate and then peeled off the stamp, leaving the target nanoflakes on the substrate. The gold array was fabricated by UV lithography, Au (45 nm)/Cr (5 nm) deposition, and subsequent acetone lift-off process. Oxygen plasma treatment was carried out to remove photoresist residues.

**Sample characterization.** The morphologies and thicknesses of exfoliated VP flakes were examined by an optical microscope (BX51, Olympus) and AFM (Oxford, Asylum Research Cypher S). Raman and PL spectra were collected using a confocal Raman system (Horiba LabRAM HR Evolution) equipped with a liquid He cryostat. The samples were illuminated by lasers with wavelengths of 532 or 633 nm. The laser power was kept below 1 mW to avoid damage to VP flakes. Optical absorption and SHG measurements were carried out on a home-built micro-optical system, and the SHG signal was excited by an ultrafast Nd: YAG laser.

**Data Availability**

The data that support the findings of this study are available from the corresponding authors upon reasonable request.

**Acknowledgments**

This work was supported by the Ministry of Science and Technology of China (Grants No. 2018YFE0202700 and No. 2021YFA1400502), the National Natural Science Foundation of China (Grants No. 11974391, No. 11825405, No. 1192780039, and No. U2032204), the

International Partnership Program of Chinese Academy of Sciences (Grant No. 112111KYSB20200012), and the Strategic Priority Research Program of Chinese Academy of Sciences (Grants No. XDB33030100 and No. XDB30000000).

**Competing Interests**

The authors declare no competing financial or non-financial interests.

**Author Contributions**

B.F., K.W., and Z.S. conceived the project. X.Z. and J.Z. synthesized the VP crystals. Z.S. and Z.C performed the experiments. Z.S., P.C., L.C., K.W., and B.F. analyzed the data. Z.S. and B.F. wrote the manuscript with comments from all authors.

**References**

[1] Liu, C. et al. Two-dimensional materials for next-generation computing technologies. *Nat. Nanotechnol.* **15**, 545–557 (2020).

[2] Huang, X.; Liu, C.; Zhou, P. 2D semiconductors for specific electronic applications: From device to system. *npj 2D Mater. Appl.* **6**, 51 (2022).

[3] Kang, S. et al. 2D semiconducting materials for electronic and optoelectronic applications: Potential and challenge. *2D Mater.* **7**, 022003 (2020).

[4] Liang, G. et al. Mid-infrared photonics and optoelectronics in 2D materials. *Mater. Today* **51**, 294–316 (2021).

[5] Ping, Y. et al. Polarization sensitive solar-blind ultraviolet photodetectors based on ultrawide bandgap $KNb_3O_8$ nanobelt with fringe-like atomic lattice. *Adv. Func. Mater.* **32**, 2111673 (2022).

[6] Schusteritsch, G.; Uhrin, M.; Pickard, C. J. Single-layered Hittorf's phosphorus: A wide-bandgap high mobility 2D material. *Nano Lett.* **16**, 2975–2980 (2016).

[7] Zhang, L. et al. High yield synthesis of violet phosphorus crystals. *Chem. Mater.* **32**, 7363–7369 (2020).

[8] Zhang, L. et al. Structure and properties of violet phosphorus and its phosphorene exfoliation. *Angew. Chem.* **132**, 1090–1096 (2020).

[9] Lin, S. et al. Liquid-phase exfoliation of violet phosphorus for electronic applications.

*SmartMat* **2**, 226–233 (2021).

[10] Li, Y. et al. Impurity-induced robust trionic effect in layered violet phosphorus. *Adv. Opt. Mater.* **10**, 2101538 (2022).

[11] Zhang, L. et al. Fast identification of the crystallographic orientation of violet phosphorus nanoflakes with preferred in-plane cleavage edge orientation. *Adv. Func. Mater.* **32**, 2111057 (2022).

[12] Ricciardulli, A. G.; Wang, Y.; Yang, S.; Samorì, P. Two-dimensional violet phosphorus: A p-type semiconductor for (opto)electronics. *J. Am. Chem. Soc.* **144**, 3660–3666 (2022).

[13] Butler, S. Z. et al. Progress, challenges, and opportunities in two-dimensional materials beyond graphene. *ACS Nano* **7**, 2898–2926 (2013).

[14] Wang, D.; Li, X. B.; Han, D.; Tian, W. Q.; Sun, H. B. Engineering two-dimensional electronics by semiconductor defects. *Nano Today* **16**, 30–45 (2017).

[15] He, Y. M. et al. Single quantum emitters in monolayer semiconductors. *Nat. Nanotechnol.* **10**, 497–502 (2015).

[16] Xu, X. et al. Localized state effect and exciton dynamics for monolayer $WS_2$. *Opt. Express* **29**, 5856 (2021).

[17] Aghajanian, M. et al. Resonant and bound states of charged defects in two-dimensional semiconductors. *Phys. Rev. B* **101**, 081201 (2020).

[18] Chaves, A. et al. Bandgap engineering of two-dimensional semiconductor materials. *npj 2D Mater. Appl.* **4**, 29 (2020).

[19] Splendiani, A. et al. Emerging photoluminescence in monolayer $MoS_2$. *Nano Lett.* **10**, 1271–1275 (2010).

[20] Bandurin, D. A. et al. High electron mobility, quantum hall effect and anomalous optical response in atomically thin InSe. *Nat. Nanotechnol.* **12**, 223–227 (2017).

[21] Li, L. et al. Direct observation of the layer-dependent electronic structure in phosphorene. *Nat. Nanotechnol.* **12**, 21–25 (2017).

[22] He, K.; Poole, C.; Mak, K. F.; Shan, J. Experimental demonstration of continuous electronic structure tuning via strain in atomically thin $MoS_2$. *Nano Lett.* **13**, 2931–2936 (2013).

[23] Li, Z. et al. Efficient strain modulation of 2D materials via polymer encapsulation. *Nat. Commun.* **11**, 1151 (2020).

[24] Zhang, L. et al. Phonon properties of bulk violet phosphorus single crystals: temperature and pressure evolution. *ACS Appl. Electron. Mater.* **3**, 1043–1049 (2021).

[25] Wang, F. et al. Honeycomb $RhI_3$ flakes with high environmental stability for

optoelectronics. *Adv. Mater.* **32**, 2001979 (2020).

[26] Shi, J. et al. 3R $MoS_2$ with broken inversion symmetry: A promising ultrathin nonlinear optical device. *Adv. Mater.* **29**, 1701486 (2017).

[27] Zhang, S. et al. Defect structure of localized excitons in a $WSe_2$ monolayer. *Phys. Rev. Lett.* **119**, 046101 (2017).

[28] Shi, J. et al. Identification of high-temperature exciton states and their phase-dependent trapping behaviour in lead halide perovskites. *Energy Environ. Sci.* **11**, 1460–1469 (2018).

[29] Saigal, N.; Ghosh, S. Evidence for two distinct defect related luminescence features in monolayer $MoS_2$. *Appl. Phys. Lett.* **109**, 122105 (2016).

[30] Wei, C. et al. Bound exciton and free exciton states in GaSe thin slab. *Sci. Rep.* **6**, 33890 (2016).

[31] Darlington, T. P. et al. Imaging strain-localized excitons in nanoscale bubbles of monolayer $WSe_2$ at room temperature. *Nat. Nanotechnol.* **15,** 854–860 (2020).

[32] Chowdhury, T. et al. Anomalous room-temperature photoluminescence from nanostrained $MoSe_2$ monolayers. *ACS Photonics* **8,** 2220–2226 (2021).

[33] Shabani, S. et al. Ultralocalized optoelectronic properties of nanobubbles in 2D semiconductors. *Nano Lett.* **22,** 7401–7407 (2022).

[34] Kim, G. et al. High-density, localized quantum emitters in strained 2D semiconductors. *ACS Nano* **16,** 9651–9659 (2022).

[35] Schmidt, T.; Lischka, K.; Zulehner, W. Excitation-power dependence of the near-band-edge photoluminescence of semiconductors. *Phys. Rev. B* **45**, 8989–8994 (1992).

[36] Greben, K.; Arora, S.; Harats, M. G.; Bolotin, K. I. Intrinsic and extrinsic defect-related excitons in TMDCs. *Nano Lett.* **20**, 2544–2550 (2020).

[37] Yan, T.; Qiao, X.; Liu, X.; Tan, P.; Zhang, X. Photoluminescence properties and exciton dynamics in monolayer $WSe_2$. *Appl. Phys. Lett.* **105**, 101901 (2014).

[38] Wu, Z. et al. Defects as a factor limiting carrier mobility in $WSe_2$: A spectroscopic investigation. *Nano Res.* **9**, 3622–3631 (2016).

[39] Wu, Z. et al. Defect activated photoluminescence in $WSe_2$ monolayer. *J. Phys. Chem. C* **121**, 12294–12299 (2017).

[40] Mak, K. F. et al. Measurement of the optical conductivity of graphene. *Phys. Rev. Lett.* **101**, 196405 (2008).

[41] Tauc, J.; Grigorovici, R.; Vancu, A. Optical properties and electronic structure of amorphous germanium. *Phys. Status Solidi (B)* **15**, 627–637 (1966).

[42] Aslan, O. B.; Deng, M.; Heinz, T. F. Strain tuning of excitons in monolayer $WSe_2$. *Phys.*

*Rev. B* **98**, 115308 (2018).

[43] Zhang, J. et al. Giant bandgap engineering in two-dimensional ferroelectric α-In$_2$Se$_3$. *J. Phys. Chem. Lett.* **13**, 3261–3268 (2022).

[44] Wang, F. et al. Anisotropic infrared response and orientation-dependent strain-tuning of the electronic structure in Nb$_2$SiTe$_4$. *ACS Nano* **16**, 8107–8115 (2022).

[45] Spirito, D. et al. Tailoring photoluminescence by strain-engineering in layered perovskite flakes. *Nano Lett.* **22**, 4153–4160 (2022).

**Figure Legends**

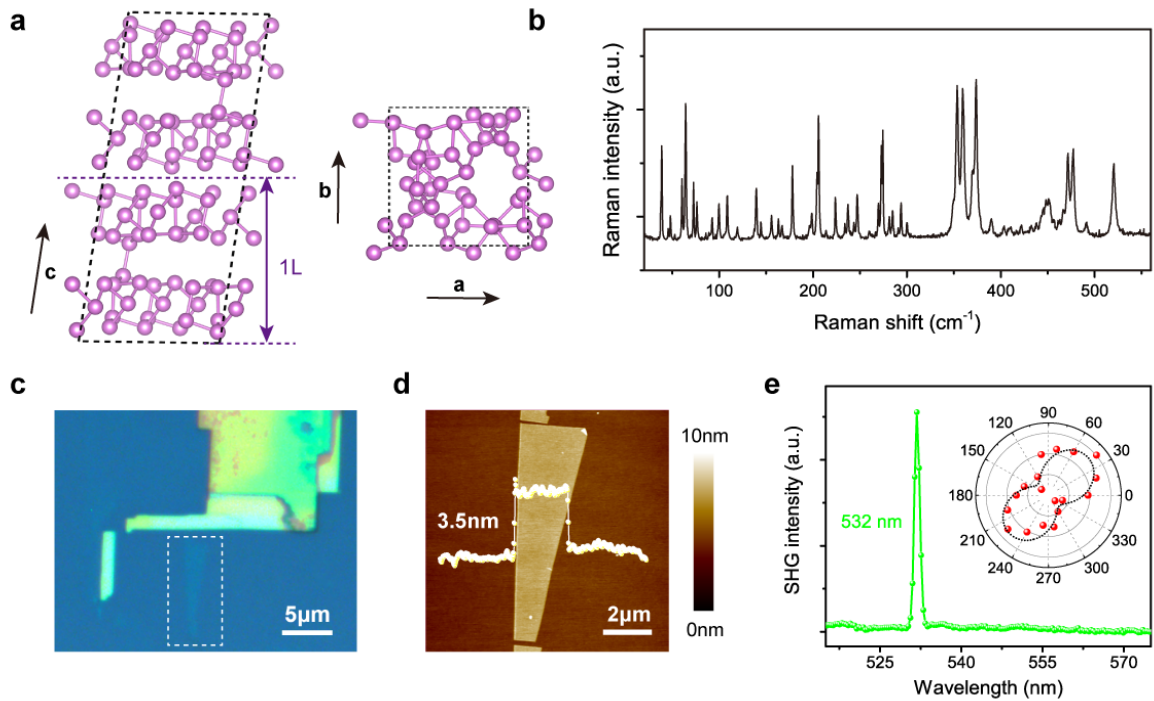

**Fig. 1 Basic characterizations of violet phosphorus.** (a) Schematic drawing of the atomic structure of VP. (b) Raman spectrum of VP. (c,d) Optical and atomic force microscopy (AFM) image of a few-layer VP nanoflake obtained by mechanical exfoliation, respectively. (e) SHG spectrum of a thick VP flake with 1064-nm ultrafast laser excitation. Inset: Polarization dependence of the SHG intensity in the parallel (XX) polarization configuration.

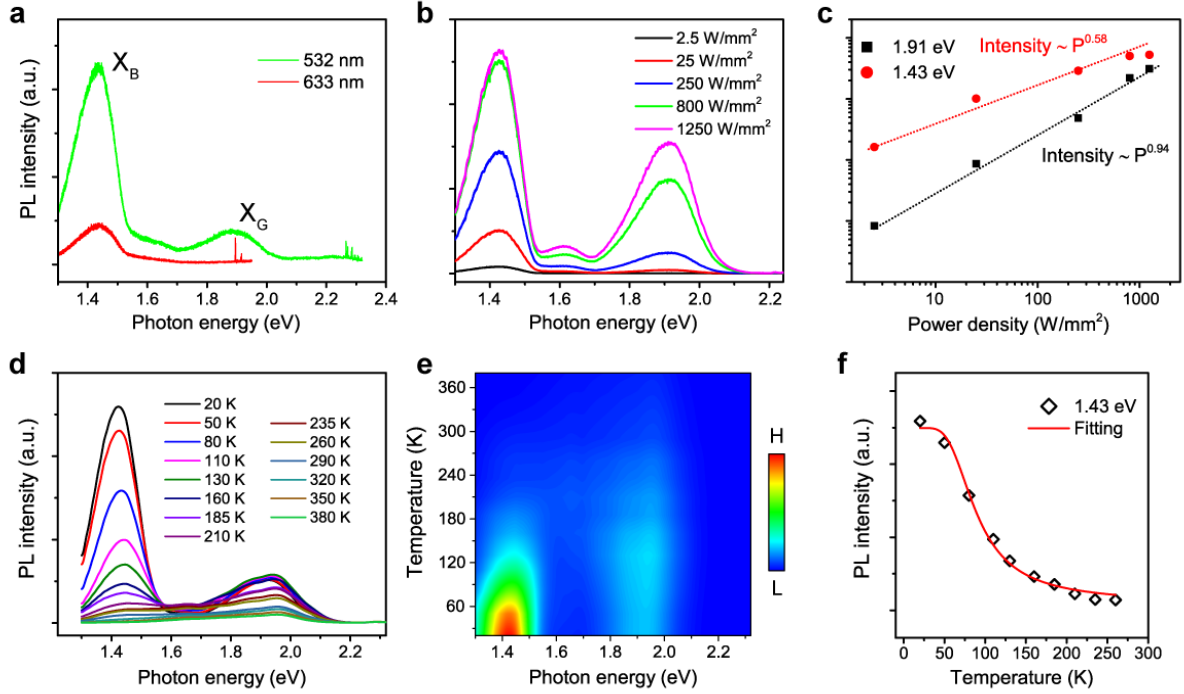

**Fig. 2 Photoluminescence spectra related to the bound excitonic emission.** (a) PL spectra of a VP nanoflake at 5 K with the 532-nm (green) and 633-nm (red) laser excitation. The thickness of the VP flake is 60 nm. (b) Excitation-power-dependent PL spectra at 5 K with 532-nm illumination. (c) Power density dependence of the 1.43-eV and 1.91-eV PL peaks extracted from (b). (d,e) Temperature-dependent PL spectra ($\lambda_{ex}$=532 nm) from 20 K to 380 K. (f) Temperature dependence of the PL intensity at 1.43 eV extracted from (d). The red line is the fitted result using the Arrhenius plot relation.

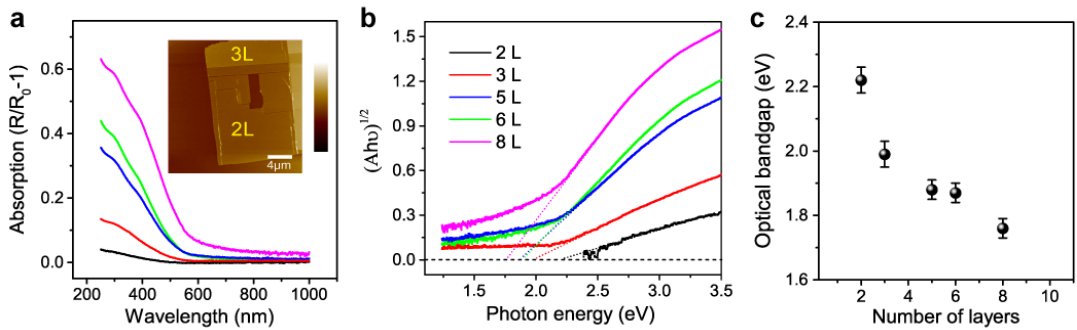

**Fig. 3 Layer-dependent bandgap evolution.** (a) Optical absorption spectra of 2D VP nanoflakes with different thicknesses. Inset: AFM image of the measured 2L and 3L area. (b) Tauc plot of the absorbance data to extract the optical bandgap. (c) Optical bandgap versus the number of layers. All error bars are determined by multiple measurements.

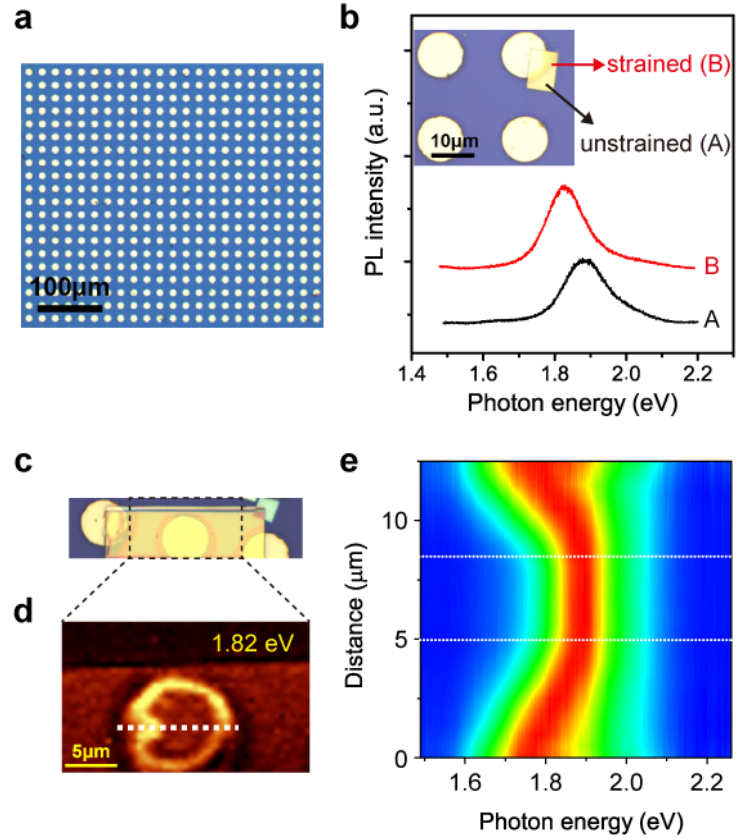

**Fig. 4 Bandgap variation under the external tensile strain.** (a) Optical image of a periodic array with circular Au disks. The diameter and height of each Au disk are 10 μm and 50 nm, respectively. The period of the array is 20 μm. (b) PL spectra of a VP nanoflake acquired on the strained and unstrained area, respectively, as indicated in the optical image in the inset. (c) Optical image that shows a VP flake which completely covers a Au disk. (d) PL mapping at the same area of (c) focused on 1.82 eV. (e) Spatial resolved normalized PL spectra along the white dashed line in (d).